\documentclass[sigconf,natbib=false,nonacm]{acmart}
\AtBeginDocument{%
  }
\usepackage{adjustbox}
\newcommand{\downg}[1]{\textcolor{green!60!black}{$\downarrow$~#1}}
\usepackage{float}
\RequirePackage[
  datamodel=acmdatamodel,
  style=acmnumeric,
  ]{biblatex}

\begin{document}

\title{SymNetPro: LOS-Aware Directional Multi-Transmitter Localization from Sparse Radio Observations}
\subtitle{Extended Preprint}

\author{Lyuzhou Ye}

\affiliation{%
  \institution{University of North Texas}
  \city{Denton}
  \state{Texas}
  \country{USA}
}
\email{lyuzhouye@my.unt.edu}
\author{Heng Fan}
\affiliation{%
  \institution{University of North Texas}
  \city{Denton}
  \state{Texas}
  \country{USA}}
\email{heng.fan@unt.edu}

\author{Yan Huang}
\affiliation{%
  \institution{University of North Texas}
  \city{Denton}
  \state{Texas}
  \country{USA}}
\email{yan.huang@unt.edu}

\begin{abstract}
Directional multi-transmitter localization from sparse received-power observations is difficult because the receiver observes only the source-unresolved aggregate field: multiple directional sources superpose, building blockage fragments their visible regions, and stronger sources can mask weaker ones. We present SymNetPro, which retains the dual-task radio-map reconstruction and localization backbone of SymNet and adds two targeted components. First, a sparse line-of-sight (LOS)-aware attention bias injects obstruction-aware spatial relations into selected token interactions. Second, transmitter-drop augmentation recomposes training scenes after removing one sample-supported transmitter, exposing the model to controlled source-cardinality variation. Experiments on directional ray-traced urban environments show substantially lower OSPA than representative localization baselines under extreme sparse sampling, with consistent gains under measurement noise and increasing transmitter count. A transmitter-specific evidence analysis further shows that remaining misses concentrate in regimes where the target contributes little distinguishable power to the aggregate observation.
\end{abstract}


\maketitle

\begin{center}
\small\emph{A shorter version of this work was accepted to ACM SIGSPATIAL 2026. This manuscript is an extended preprint and is not the conference proceedings version. The code and dataset are publicly available at \url{https://github.com/LyuzhouYe98/SymNet--a-multi-task-network-for-joint-radio-map-reconstruction-and-transmitter-localization}.}
\end{center}

\section{Introduction}
\label{sec:intro}
Wireless systems increasingly rely on directional propagation, such as mmWave communication and beamforming-based 5G networks, where received power is highly sensitive to antenna orientation, building occlusion, and reflection. We consider a passive, non-cooperative sensing setting in which only source-unresolved received-signal-strength (RSS) measurements are available. At each sampled location, the receiver observes the aggregate received power of all active transmitters, without transmitter-specific pilots, identifiers, or ranging measurements. The individual source contributions are therefore not directly separable.

This source superposition creates a distinct difficulty for directional multi-transmitter localization. At a sampled location, let $S_m$ be the linear received power of a target transmitter and let $I_m$ be the sum of the linear powers from all other active transmitters. We define the linear signal-to-interference ratio (SIR) as $\rho_m=S_m/I_m$ and its dB form as $\mathrm{SIR}_m^{\mathrm{dB}}=10\log_{10}\rho_m$. Relative to the interference-only field, adding the target changes the aggregate dB measurement by $\Delta P_m=10\log_{10}(1+\rho_m)$. Therefore, $\mathrm{SIR}=-10$~dB produces only a $0.41$~dB change, while $-20$~dB produces only about $0.043$~dB. A transmitter can thus be physically present yet leave little distinguishable evidence in the aggregate RSS field. Directional beams and building blockage further make source-specific evidence spatially fragmented.

Existing radio map learning and transmitter localization methods have made substantial progress in single-transmitter or omnidirectional settings~\cite{DCNet,dsloc,tldl,LocNet,ACTGAN}. Many learning-based approaches treat sparse signal observations as image-like inputs and apply convolutional, reconstruction-based, heatmap-based, or detection-based models. In directional multi-transmitter scenes, however, individual sources need not correspond to separable visual patterns: weaker sources can be masked by stronger ones, and the evidence supporting each transmitter can be unevenly distributed across sparse samples.

To address these challenges, we propose \textbf{SymNetPro}, a LOS-aware framework for sparse directional radio map learning and multi-transmitter localization. We build on the ViT-based dual-task reconstruction/localization backbone of SymNet~\cite{SymNet} and focus on two targeted extensions rather than backbone scaling. First, we introduce a sparse LOS-aware patch bias that guides attention toward propagation-consistent relations. Second, we propose transmitter-drop augmentation, which constructs counterfactual training views by removing one sample-supported transmitter and recomposing both the radio map and localization target. The auxiliary reconstruction branch is inherited from SymNet; our contribution is the adaptation to source-unresolved directional multi-source recovery.

In addition to improving localization performance, we analyze transmitter-specific exclusive evidence as a diagnostic factor for missed detections. We define exclusive evidence using a target-versus-strongest-competitor dB margin and relate this pairwise dominance criterion to aggregate signal-to-interference ratio (SIR) and local sensitivity. The analysis is post-hoc and is not provided to the model at inference.

Our contributions are summarized as follows. \textbf{First}, we study passive, source-unresolved directional multi-transmitter localization from sparse RSS with auxiliary radio map reconstruction and show substantially lower OSPA than strong baselines in the target sparse regime. \textbf{Second}, we introduce a parameter-light sparse LOS-aware patch bias that injects obstruction-aware spatial relations into token interactions. \textbf{Third}, we propose transmitter-drop augmentation for source-cardinality robustness using training-only source components. \textbf{Finally}, we provide a transmitter-specific evidence analysis that connects pairwise source dominance, aggregate SIR, and localization failures under sparse superposition.

\section{Related Work}
\subsection{Radio Map Reconstruction}

Radio map construction aims to infer dense received-power fields from sparse measurements and environmental information. Many transmitter localization algorithms are based on radio map reconstruction. Traditional approaches include parametric methods, such as compressed sensing~\cite{Compressed_sensing} and dictionary learning~\cite{dictionary}, and non-parametric interpolation methods, such as Kriging~\cite{Oridinary_Kriging} and radial basis function interpolation~\cite{RadialBasic}. These methods often rely on propagation assumptions or prior knowledge of the transmitter configuration, which can be difficult to satisfy in dense urban environments with blockage and multipath effects.

Deep learning methods have become strong alternatives. RadioUNet~\cite{Levie2019RadioUNetFR} established a U-Net-based formulation for radio map estimation, and later models improved reconstruction through stronger encoder--decoder networks, generative modeling, spatial pyramid pooling, and cascaded refinement. Representative methods include DeepAE~\cite{DEEPAE}, SkipNet~\cite{SkipNet}, GAN-CRME~\cite{CGAN}, ACT-GAN~\cite{ACTGAN}, and DC-Net~\cite{DCNet}. More recently, ViT-RefineNet~\cite{ViTRefineNet} and SymNet~\cite{SymNet} use attention-based backbones to capture long-range spatial dependencies and provide strong baselines for directional radio map reconstruction.

Most existing models incorporate environmental structure through input channels such as building maps or distance-to-building features~\cite{SymNet}. These inputs provide useful local geometric context, but they do not explicitly encode pairwise line-of-sight relations between spatial regions during token interaction. In this work, our primary focus is multi-transmitter localization under sparse directional observations.
\subsection{Transmitter Localization}

Traditional transmitter localization methods, such as time of arrival (TOA), time difference of arrival (TDOA), and angle of arrival (AOA)~\cite{gezici2008survey,gustafsson2005mobile}, often require dedicated hardware support and can be difficult to deploy at scale. RSS-based localization is easier to implement~\cite{832252,patwari2005locating}, but it is sensitive to non-line-of-sight propagation and often depends on propagation models that become unreliable in dense urban environments~\cite{haeberlen2004practical}.

Deep learning methods have therefore been explored to learn localization cues directly from measurements. Coordinate-based methods predict transmitter locations explicitly. For example, Zhang et al.~\cite{zhang} combined an MLP feature extractor with an HMM to predict transmitter coordinates. DeepTxFinder~\cite{zubow2020deeptxfinder} maps sparse RSS readings into grid cells and uses a CNN--MLP design to estimate the number and positions of transmitters. MT-GCNN~\cite{wang} further performs containment-cell classification and distance regression in a multi-task gated CNN framework. These methods are effective in certain settings, but can be sensitive to sensor layout, sampling density, and environmental changes.

Heatmap-based formulations have become common alternatives. TL;DL~\cite{tldl} predicts transmitter heatmaps with a U-Net and extracts coordinates through thresholding and peak suppression. LocNet~\cite{LocNet} shows that a lightweight U-Net-style model can achieve competitive localization performance with fewer parameters. DSLoc~\cite{dsloc} addresses sparse localization by introducing bias correction, sparse expansion preprocessing, an HRNet backbone, and centroid regression for finer localization. These methods show the effectiveness of heatmap-based localization, but many are developed under omnidirectional or simplified propagation assumptions, where transmitter locations are closely related to strong signal regions.

In directional multi-transmitter settings, this relationship becomes less reliable. Transmitter orientation and building blockage can shift the strongest received regions away from the transmitter, and superposition can cause weaker transmitters to be masked by stronger sources. Therefore, individual transmitters may not appear as separable visual objects in the sparse input, and localization performance depends on whether each transmitter has sufficient source-specific evidence.

Multi-transmitter localization can be approached through heatmap prediction, object detection, or localization from reconstructed radio maps. DeepMTL~\cite{deepmtl} follows a detection-based pipeline: it first predicts a Gaussian-like transmitter target map and then applies YOLOv3~\cite{yolo} to detect transmitter locations from the predicted target representation. This formulation converts multi-transmitter localization into an object detection problem, but its performance depends on both the quality of the intermediate transmitter-target map and the detector.

Another line of work first reconstructs the dense radio map and then infers transmitter positions from the predicted field. For example, Obadah et al.~\cite{cGANM} use persistent-homology-based analysis on the predicted radio map to locate multiple transmitters. LRM-MSL~\cite{LRMMSL} also targets multiple-source localization by reconstructing the radio map and then applying a trained localization module to infer multiple transmitter positions.

Compared with single-transmitter localization, multi-transmitter recovery is substantially more challenging. The number of active transmitters can vary, and the observed signal is a superposition of multiple source contributions. A sampled location may be dominated by one transmitter while providing weak evidence for another, which makes missed detections and false alarms more likely. SymNet~\cite{SymNet} is the closest prior work to our setting because it jointly considers radio map reconstruction and transmitter localization from sparse observations. Our work extends this direction by introducing LOS-aware token interactions, transmitter-drop augmentation, and a transmitter-specific evidence analysis for understanding missed detections under directional multi-source superposition.
\section{Preliminary}

\subsection{Problem Definition}

Given an ROI discretized into an $H\times W$ grid, a scene contains $N$ active directional transmitters indexed by $\mathcal{T}=\{1,\ldots,N\}$, where $N$ and the active set are not given to the model at inference. Transmitter $m$ is parameterized by
\[
\psi_m=(x_m,y_m,\phi_m),
\]
where $(x_m,y_m)$ is its location and $\phi_m$ its antenna orientation. Let $P_m(i,j)$ denote its single-transmitter received power in dBm at cell $(i,j)$. The propagation maps use $P_{\min}=-120$~dBm as the no-signal cutoff. We represent each valid contribution in linear power (mW) as
\begin{equation}
S_m(i,j)=
\begin{cases}
10^{P_m(i,j)/10}, & P_m(i,j)>P_{\min},\\
0, & P_m(i,j)\le P_{\min}.
\end{cases}
\end{equation}
The source-unresolved aggregate power is
\begin{equation}
S_{\mathrm{agg}}(i,j)=\sum_{m\in\mathcal{T}}S_m(i,j),
\end{equation}
and the corresponding aggregate dB map is
\begin{equation}
P_{\mathrm{agg}}(i,j)=
\begin{cases}
10\log_{10}S_{\mathrm{agg}}(i,j), & S_{\mathrm{agg}}(i,j)>0,\\
P_{\min}, & S_{\mathrm{agg}}(i,j)=0.
\end{cases}
\end{equation}
Thus, the receiver observes only the total power contributed by all active transmitters at a location, not the individual $P_m$ or $S_m$ components.

We apply a fixed dataset-wide normalization
\begin{equation}
Y(i,j)=\frac{P_{\mathrm{agg}}(i,j)-P_{\min}}{P_{\max}-P_{\min}},
\qquad P_{\max}=-5.041252~\mathrm{dBm},
\end{equation}
so the no-signal cutoff maps to zero. No per-map normalization is used. Only locations in the sampling set $\Omega$ are observed:
\begin{equation}
R(i,j)=
\begin{cases}
Y(i,j),&(i,j)\in\Omega,\\
0,&\text{otherwise}.
\end{cases}
\end{equation}
We encode the sparse observation and environment as a two-channel input
\begin{equation}
\mathbf{X}=\mathrm{Concat}(R,B_\Omega),
\end{equation}
where
\[
b_{i,j}=
\begin{cases}
-1,&(i,j)\text{ is a building pixel},\\
1,&(i,j)\in\Omega\text{ and is not a building pixel},\\
0,&\text{otherwise}.
\end{cases}
\]
This mask distinguishes an unobserved zero from a sampled non-building cell.

For later analysis, when $N\ge2$ we define the aggregate interference seen relative to target transmitter $m$ as
\begin{equation}
I_m(i,j)=\sum_{q\in\mathcal{T}\setminus\{m\}}S_q(i,j),
\end{equation}
the linear SIR as $\rho_m=S_m/I_m$, and its dB form as
\begin{equation}
\mathrm{SIR}_m^{\mathrm{dB}}(i,j)=10\log_{10}\rho_m(i,j).
\end{equation}
The target's fractional contribution to the aggregate power is
\begin{equation}
\alpha_m(i,j)=\frac{S_m(i,j)}{S_m(i,j)+I_m(i,j)}
=\frac{\rho_m(i,j)}{1+\rho_m(i,j)}.
\end{equation}
These quantities are analysis variables computed from the known source components; they are not inference inputs.

For target construction, define the fixed-normalized single-source component
\begin{equation}
\widetilde{Y}_m(i,j)=
\begin{cases}
\dfrac{P_m(i,j)-P_{\min}}{P_{\max}-P_{\min}}, & P_m(i,j)>P_{\min},\\
0, & P_m(i,j)\le P_{\min}.
\end{cases}
\end{equation}
A transmitter is called \emph{sample-supported} if this component is nonzero at least once among the sampled locations:
\begin{equation}
\mathcal{M}_{\mathrm{sup}}
=
\left\{m\in\mathcal{T}:\exists(i,j)\in\Omega\ \text{s.t.}\ \widetilde{Y}_m(i,j)>0\right\}.
\end{equation}
Only transmitters in $\mathcal{M}_{\mathrm{sup}}$ are included in the localization target. Source-specific component maps are used only to construct training/evaluation targets and the post-hoc evidence analysis. At inference, the model receives only the aggregate observation $R$ and $B_\Omega$, not individual source maps, source identities, orientations, $N$, SIR, or $\alpha_m$. The task is to predict the dense aggregate radio map and recover the locations and cardinality of the sample-supported transmitter set.

\section{Method}

\subsection{Overview}
\begin{figure*}
    \centering
    \includegraphics[width=1\linewidth]{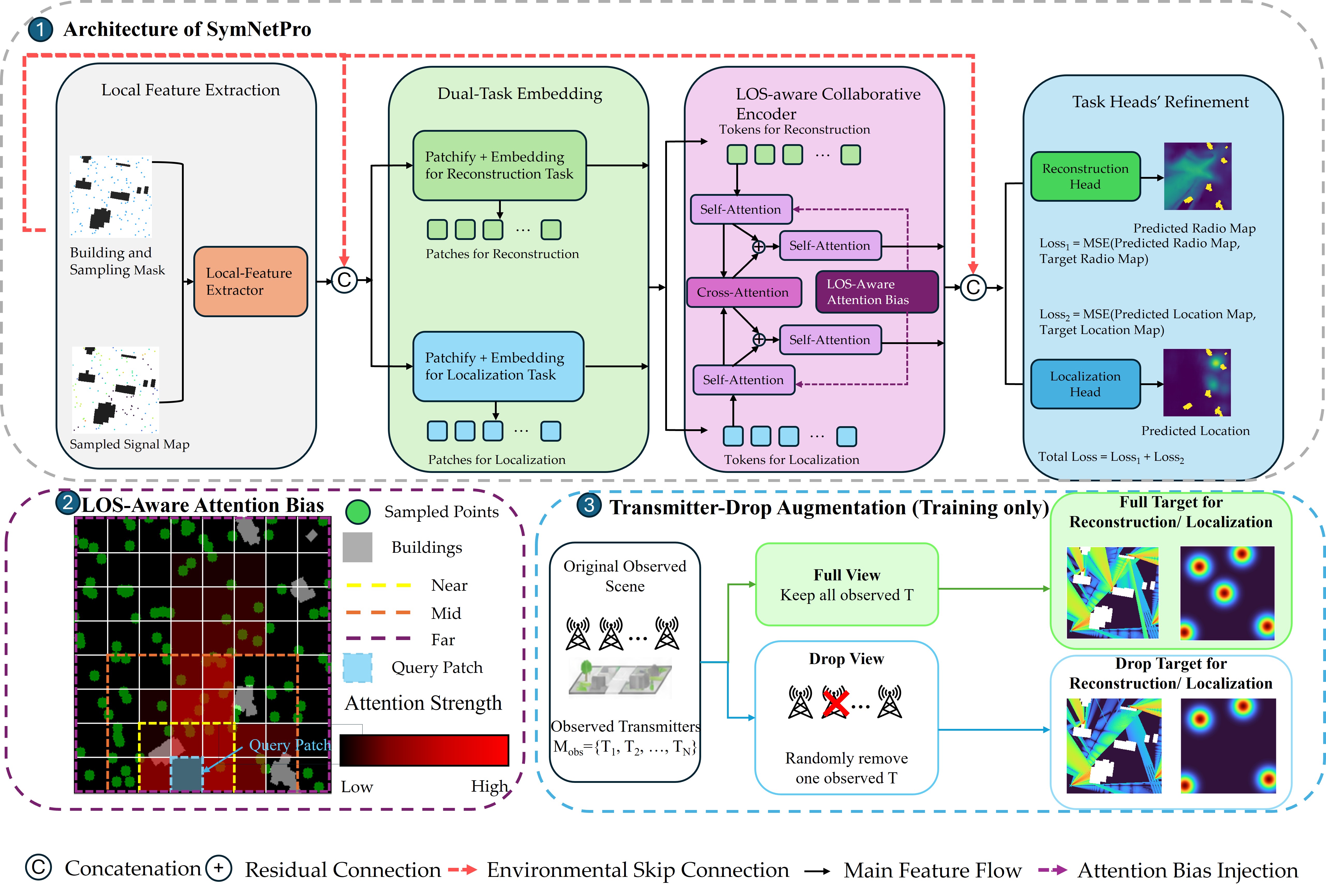}
    \Description{Architecture of SymNetPro with the inherited dual-task backbone, sparse LOS-aware attention bias, and transmitter-drop training augmentation.}
    \caption{SymNetPro retains the SymNet dual-task backbone and adds a sparse LOS-aware attention bias and transmitter-drop augmentation. Source-specific components are used only to construct training views and are not inference inputs.}
    \label{fig1}
\end{figure*}
We build on the ViT-based dual-task backbone introduced in SymNet~\cite{SymNet}. The model receives the two-channel tensor $\mathbf{X}=\mathrm{Concat}(R,B_\Omega)$ and jointly predicts a dense aggregate radio map and a Gaussian transmitter-location heatmap. The inherited backbone contains a local feature extractor, task-specific patch embeddings, collaborative attention, and task-specific refinement heads. Compared with the prior work, the DNB channel is removed.

Our methodological changes are limited to two components designed for directional multi-source recovery: (i) a sparse LOS-aware attention prior that modifies selected pairwise token preferences while preserving the dense self-attention backbone, and (ii) transmitter-drop augmentation that exposes the model to controlled source-cardinality changes during training.

\subsection{LOS-aware Attention Bias}

Directional radio propagation is strongly affected by distance and building blockage. We therefore construct a sparse patch-level geometric prior in two distinct stages: candidate-graph construction selects propagation-relevant patch pairs, and a separate LOS-aware score constructs the attention bias on the selected edges.

\paragraph{Patch-level representation.}
The $256\times256$ grid is divided into non-overlapping $8\times8$-pixel patches, yielding $32\times32=1024$ tokens. Each valid patch uses its center as a representative point, or the nearest valid pixel to the center when the center is blocked. For source patch $i$, let $n_i$ denote the number of sampled valid pixels in the patch and define the sampling confidence
\begin{equation}
q_i=\min(1,n_i/4).
\end{equation}
For patches with representative points $r_i$ and $r_j$, the normalized distance is
\begin{equation}
d_{ij}=\frac{\lVert r_i-r_j\rVert_2}{\sqrt{H^2+W^2}}.
\end{equation}

\paragraph{Candidate-graph construction.}
For each target patch $j$, candidate source patches are divided into near, middle, and far ranges $[0,50)$, $[50,100)$, and $[100,\infty)$ pixels. Within each range, candidates are first ranked by
\begin{equation}
c_{ij}=\log(q_i+\epsilon)-\lambda_c d_{ij},\qquad \lambda_c=1,
\end{equation}
where $\epsilon>0$ is a numerical stabilizer. At most $(12,12,8)$ candidates are retained from the three ranges. This stratification prevents the graph from collapsing to only nearby patches. For the retained candidates, we compute the LOS-aware ranking weight
\begin{equation}
w_{ij}
=
q_i\exp\!\left[
-\frac{d_{ij}\{1+\lambda_{\mathrm{los}}(1-\ell_{ij})\}}{\tau_c}
\right],
\end{equation}
where $\ell_{ij}\in\{0,1\}$ is the binary LOS indicator obtained by segment--building intersection, $\lambda_{\mathrm{los}}=1$, and $\tau_c=0.12$. Candidates are ranked by $w_{ij}$ and truncated to at most $K=32$ source patches per target; denote this final selected set by $\mathcal{N}(j)$. The temperature $\tau_c$ is used only for candidate selection and is not the attention-bias temperature below.

\paragraph{Attention-bias construction.}
For each selected edge $i\rightarrow j$, we define a separate bias score
\begin{equation}
e_{ij}=-\lambda_d d_{ij}-\lambda_{\mathrm{nlos}}(1-\ell_{ij}),
\qquad \lambda_d=2,\quad \lambda_{\mathrm{nlos}}=1.
\end{equation}
The selected scores are normalized row-wise using an independent temperature $\tau_{\mathrm{bias}}=2$:
\begin{equation}
\pi_{ij}
=
\frac{\exp(e_{ij}/\tau_{\mathrm{bias}})}
{\sum_{i'\in\mathcal{N}(j)}\exp(e_{i'j}/\tau_{\mathrm{bias}})}.
\end{equation}
We then form $B_{ji}=\log(\pi_{ij}+\epsilon)$ and center the nonzero bias values within each target row. Non-selected patch pairs receive no explicit geometric prior and retain their ordinary attention logits.

\paragraph{Injecting sparse bias into attention.}
The prior is added to the attention logits:
\begin{equation}
\mathrm{Attn}(Q,K,V)
=
\mathrm{Softmax}\!\left(
\frac{QK^\top}{\sqrt{d_k}}+\gamma B
\right)V,
\end{equation}
where the learnable scale $\gamma$ is initialized to $0.01$. With $T=1024$ tokens and embedding width $d$, the main self-attention remains dense with $O(T^2d)$ complexity, while only $O(TK)$ pairwise geometric priors are retained. Thus, the proposed module is parameter-light but introduces additional graph/bias construction cost, which we quantify in the runtime analysis.

\subsection{Transmitter-Drop Augmentation}

Source-specific component maps are available while constructing the training data because each multi-transmitter scene is synthesized from known single-transmitter maps. These source identities and component maps are privileged training information and are never provided to the network at inference.

For a sampling mask $\Omega$, let $\mathcal{M}_{\mathrm{sup}}$ be the sample-supported set defined in the problem formulation. The full training view is recomposed from this set:
\begin{equation}
\mathcal{M}_{\mathrm{full}}=\mathcal{M}_{\mathrm{sup}},
\qquad
S_{\mathrm{full}}(i,j)=\sum_{m\in\mathcal{M}_{\mathrm{full}}}S_m(i,j).
\end{equation}
When $|\mathcal{M}_{\mathrm{sup}}|>1$, we sample one transmitter $r$ uniformly from $\mathcal{M}_{\mathrm{sup}}$ and construct the drop view
\begin{equation}
\mathcal{M}_{\mathrm{drop}}=\mathcal{M}_{\mathrm{sup}}\setminus\{r\},
\qquad
S_{\mathrm{drop}}(i,j)=\sum_{m\in\mathcal{M}_{\mathrm{drop}}}S_m(i,j).
\end{equation}
Both views are superposed in linear power, converted back to dB, normalized with the same fixed dataset bounds, and sampled using the same mask. The Gaussian transmitter-location target is regenerated from the corresponding remaining transmitter set. Thus, the input and both task targets are changed consistently, exposing the model to controlled source-cardinality variation without changing the inference interface.

\subsection{Training Objective}

For a training view $v\in\{\mathrm{full},\mathrm{drop}\}$, the model predicts a dense radio map $\hat{Y}_v$ and a localization heatmap $\hat{A}_v$. The localization target $A_v$ is constructed by placing Gaussian peaks centered at the transmitter locations retained in view $v$, using standard deviation $\sigma=17$ pixels. We use MSE losses
\begin{equation}
\mathcal{L}_{\mathrm{map}}^{v}=\mathrm{MSE}(\hat{Y}_v,Y_v),
\qquad
\mathcal{L}_{\mathrm{tx}}^{v}=\mathrm{MSE}(\hat{A}_v,A_v).
\end{equation}
The localization-loss weight is $1/2.5$ for the full view and $1$ for the drop view:
\begin{equation}
\mathcal{L}_{\mathrm{full}}
=\mathcal{L}_{\mathrm{map}}^{\mathrm{full}}+
\frac{1}{2.5}\mathcal{L}_{\mathrm{tx}}^{\mathrm{full}},
\end{equation}
\begin{equation}
\mathcal{L}_{\mathrm{drop}}
=\mathcal{L}_{\mathrm{map}}^{\mathrm{drop}}+
\mathcal{L}_{\mathrm{tx}}^{\mathrm{drop}},
\end{equation}
with total objective
\begin{equation}
\mathcal{L}=0.25\left(\mathcal{L}_{\mathrm{full}}+\mathcal{L}_{\mathrm{drop}}\right).
\end{equation}
If $|\mathcal{M}_{\mathrm{sup}}|\le1$, the invalid drop-view loss is masked out. Training uses Adam with batch size 16, a base learning rate of $5\times10^{-4}$ and $2.5\times10^{-4}$ for the local-feature extractor and task heads. Checkpoints are saved each epoch and selected by validation performance.

\subsection{Transmitter Coordinate Extraction}

During inference, the localization branch outputs a continuous heatmap $\hat{A}\in\mathbb{R}^{H\times W}$. We retain local maxima above $\tau_{\mathrm{loc}}=0.25$ and apply non-maximum suppression with an $11\times11$ window. The remaining peaks form the predicted transmitter set $\hat{\mathcal{P}}$ and therefore determine both locations and predicted cardinality. No additional detector is trained for SymNetPro.

\section{Experiments}

We evaluate SymNetPro under two sensing protocols. In the extreme-sparse regime, all methods are trained with 100 sampled points and evaluated from 100 to 500 sampled points, providing a controlled same-budget comparison. In the denser regime, cGAN and LRM-MSL are trained with 3300 sampled points and DeepMTL with 3932, matching the sensor densities used in their original $100\times100$ settings after scaling to $256\times256$; SymNetPro remains trained with 100 points. All methods are then evaluated from 660 to 3300 sampled points. We additionally test robustness to measurement noise, source-count variation, component ablations, computational cost, and transmitter-specific evidence.

A key challenge is that the total number of sampled pixels does not fully describe how much information is available for each transmitter. We therefore use a post-hoc transmitter-level evidence measure in the failure analysis. For a multi-source scene ($N\ge2$), define the pairwise dominance margin of transmitter $m$ against its strongest active competitor as
\begin{equation}
D_m(s_k)=P_m^{\mathrm{dB}}(s_k)-\max_{q\in\mathcal{T}\setminus\{m\}}P_q^{\mathrm{dB}}(s_k).
\end{equation}
The exclusive evidence set and count are
\begin{equation}
\mathcal{E}_m(\delta)=\{s_k:D_m(s_k)>\delta\},
\qquad
N_m(\delta)=|\mathcal{E}_m(\delta)|.
\end{equation}
We use $\delta=10$~dB. This is a target-versus-strongest-competitor margin, not a 10~dB aggregate-SIR threshold. Let $a=10^{\delta/10}$. If $D_m(s_k)>\delta$, then $S_m(s_k)>aS_q(s_k)$ for every competing transmitter $q$. Hence
\begin{equation}
I_m(s_k)<\frac{N-1}{a}S_m(s_k),
\end{equation}
which gives the main-body SIR and target-fraction bounds
\begin{equation}
\mathrm{SIR}_m^{\mathrm{dB}}(s_k)>
\delta-10\log_{10}(N-1),
\qquad
\alpha_m(s_k)>
\frac{1}{1+(N-1)10^{-\delta/10}}.
\end{equation}
For $\delta=10$~dB and $N=5$, these become $\mathrm{SIR}>3.98$~dB and $\alpha_m>0.714$. Appendix~\ref{app:sensitivity} connects $\alpha_m$ to local measurement sensitivity, and Appendix~\ref{app:pairwise} gives the exact relation between the pairwise margin and conventional aggregate SIR. The exclusive count is used only for post-hoc analysis and is never provided to the model.

We study the following questions: sparse and dense localization performance, robustness to 4~dB measurement noise, sensitivity to transmitter count, component and bias-design ablations, computational efficiency, and the relationship between exclusive evidence and missed transmitters.

\subsection{Dataset}
We construct the multi-transmitter dataset directly from the single-transmitter directional propagation maps released with ViT-RefineNet~\cite{ViTRefineNet}; no new ray tracing is performed in this work. Each $256\times256$ ROI has 1~m/pixel resolution and covers $256\times256$~m. The underlying maps were generated with MATLAB RF Toolbox at 28~GHz using 2~m transmitter/receiver heights, a Gaussian phased antenna with $20^\circ$ azimuth and $10^\circ$ elevation beamwidth, one reflection, no diffraction, and a $-120$~dBm cutoff~\cite{ViTRefineNet}. Each environment contains 100 single-transmitter maps with varying transmitter locations and orientations.

For each environment and $N\in\{1,2,3,4,5\}$, we form 100 multi-transmitter configurations by randomly selecting $N$ single-source component maps and superposing their valid contributions in linear power. The train/validation/test split contains 274/59/59 building environments and is environment-level, so test buildings are unseen during training. Training and validation use 10 sampling-mask replicates per configuration, yielding
\[
274\times5\times100\times10=1{,}370{,}000
\]
training samples and
\[
59\times5\times100\times10=295{,}000
\]
validation samples. The test split contains 100 mask replicates per configuration (2.95M cases per density); the main reported evaluation uses replicates 0--49, i.e., 1.475M cases per sampling density. Sampling locations are drawn uniformly without replacement from non-building pixels, and identical evaluation masks are used across methods.

For the noise experiment, zero-mean Gaussian noise with standard deviation 4~dB is added to the sampled RSS values only at test time; models are not retrained on noisy observations.

\subsection{Evaluation Metrics}
\begin{figure*}
    \centering
    \includegraphics[width=1\linewidth]{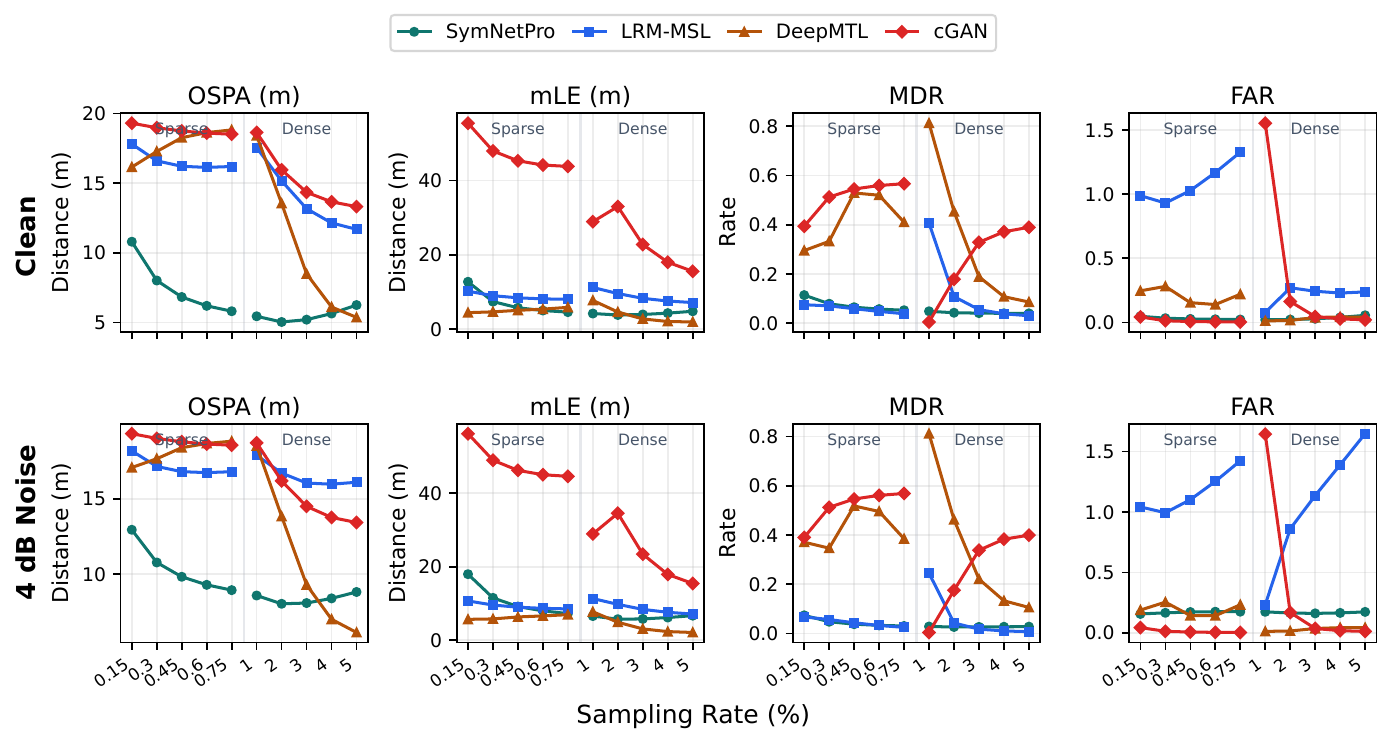}
    \Description{Localization metrics across sampling densities for SymNetPro and three baselines under clean and 4 dB noisy observations.}

    \caption{Performance in low/high sampling rates with/without noise}
    \label{fig:rq123}
\end{figure*}

For transmitter recovery, we use OSPA distance, matched localization error (mLE), missed detection rate (MDR), and false alarm rate (FAR). Let
\[
\mathcal{G}=\{g_1,\ldots,g_{N_g}\}
\]
denote the set of ground-truth transmitter locations and
\[
\mathcal{P}=\{p_1,\ldots,p_{N_p}\}
\]
denote the set of predicted transmitter locations.

OSPA~\cite{OSPA} measures the distance between two finite sets while penalizing both localization error and cardinality mismatch. For \(N_g\le N_p\), it is defined as
\[
\mathrm{OSPA}_{p,c}(\mathcal{G},\mathcal{P})
=
\left[
\frac{1}{N_p}
\left(
\min_{\pi\in\Pi_{N_p}}
\sum_{i=1}^{N_g}
d_c(g_i,p_{\pi(i)})^p
+
c^p(N_p-N_g)
\right)
\right]^{1/p}.
\]
where
\[
d_c(g,p)=\min(c,\|g-p\|_2),
\]
\(c\) is the cutoff distance, \(p\) is the order parameter, and \(\Pi_{N_p}\) denotes the set of permutations over \(N_p\) predicted locations. When \(N_g>N_p\), the definition is symmetric by swapping \(\mathcal{G}\) and \(\mathcal{P}\). In our experiments, we set \(c=20\) and \(p=2\).

To separately evaluate localization accuracy for correctly detected transmitters, we also report matched localization error (mLE). After matching predicted transmitters to ground-truth transmitters using the same distance threshold, let \(\mathcal{M}\) denote the set of matched pairs. mLE is defined as
\[
\mathrm{mLE}
=
\frac{1}{|\mathcal{M}|}
\sum_{(g,p)\in\mathcal{M}}
\lVert g-p\rVert_2 .
\]
If no transmitter is matched, mLE is not computed for that sample and is excluded from the average.

We further report missed detection rate (MDR) and false alarm rate (FAR) following the benchmark definitions. We use cardinality-only MDR and FAR. Both are normalized by the number of ground-truth transmitters:
\[
\mathrm{MDR}
=
\frac{\max(N_g-N_p,0)}{N_g},
\qquad
\mathrm{FAR}
=
\frac{\max(N_p-N_g,0)}{N_g}.
\]
MDR measures under-estimation of the transmitter count, while FAR measures over-estimation normalized by the number of true transmitters. Unlike mLE and OSPA, MDR and FAR do not depend on localization distance.

\subsection{Comparison Models}

We compare our method with representative baselines for multi-transmitter recovery. 
We evaluate methods that can produce transmitter estimates either directly or through post-processing of the predicted radio map:
\begin{itemize}
    \item \textbf{cGAN+PH}~\cite{cGANM}: cGAN reconstructs the radio map and persistent homology extracts transmitter candidates.
    \item \textbf{DeepMTL+YOLO-V3}~\cite{deepmtl}: DeepMTL predicts an intermediate transmitter-target map and YOLO-V3 recovers transmitter locations.
    \item \textbf{LRM-MSL+SourceNet}~\cite{LRMMSL}: LRM-MSL reconstructs the radio map and SourceNet recovers multiple transmitter locations.
\end{itemize}
All three baselines are retrained on the same environment split. Method-specific post-processing thresholds are selected by validation OSPA: 0.50 for cGAN+PH, 0.80/0.90 for DeepMTL+YOLO-V3 in the sparse/dense protocols, and 127/190 on the 0--255 first-stage map for LRM-MSL+SourceNet in the sparse/dense protocols.
\subsection{RQ1: Performance under Extreme Sparse Sampling Settings}
We evaluate all methods from 100 to 500 sampled points after training every method with the same 100-sample budget. As shown in Figure~\ref{fig:rq123}, SymNetPro consistently achieves the lowest OSPA in this regime, decreasing from 10.84~m at 100 sampled points to 5.81~m at 500. The baselines exhibit different cardinality failure modes: cGAN tends to under-detect sources, whereas LRM-MSL produces many false alarms; DeepMTL is more competitive at 100 points but remains substantially worse than SymNetPro and degrades when evaluated at larger sparse budgets despite being trained at 100 points.

The failure modes are especially clear for five-source scenes at 100 samples. SymNetPro misses 0.878 transmitters and produces 0.101 false detections on average (MDR/FAR $=0.176/0.020$). DeepMTL misses 1.735 transmitters (MDR $=0.347$), while LRM-MSL produces 4.091 false detections for five true sources (FAR $=0.818$). cGAN exhibits the opposite failure mode, missing 2.759 sources on average. These results show that sparse-regime performance is limited not only by coordinate accuracy but also by reliable source-cardinality recovery. Qualitative examples are provided in Figure~\ref{figf} of Appendix~\ref{app:theory}.

\subsection{RQ2: Robustness under Standard Sampling Density}
For the denser protocol, cGAN and LRM-MSL are trained with 3300 sampled points and DeepMTL with 3932, matching their original sensor densities after scaling to the $256\times256$ grid. SymNetPro remains trained with only 100 samples. All methods are evaluated from 660 to 3300 sampled points (approximately 1--5\% of the grid).

SymNetPro remains strongest from 660 to 2640 sampled points, despite its substantially lower training sensing budget. DeepMTL improves as the sampling density increases and obtains the best OSPA at 3300 points, showing that it benefits strongly from dense observations. LRM-MSL and cGAN remain less competitive, particularly in cardinality control. Overall, the results show that SymNetPro retains strong localization performance as measurement density increases while being trained only in the sparse regime. Qualitative examples are provided in Figure~\ref{figf}.

\subsection{RQ3: Robustness under Random Measurement Noise}
We add zero-mean Gaussian noise with standard deviation 4~dB to the sampled RSS values at test time without retraining. As shown in Figure~\ref{fig:rq123}, SymNetPro remains robust across sparse and denser sensing regimes. At 100 sampled points, its average OSPA is about 13.0~m across source counts; performance improves to roughly 8--9~m as sampling increases. The experiment therefore measures robustness to corrupted measurements rather than adaptation through noisy-data training.

Noise mainly increases false alarms relative to the clean setting, consistent with perturbations creating additional local responses in the predicted heatmap. Nevertheless, the method preserves substantially useful localization performance under 4~dB measurement uncertainty.

\subsection{RQ4: Performance over Different Number of Transmitters}
\begin{table}[t]
    \centering
    \caption{Per-source-count localization at 3\% sampling (1980 points). MDR and FAR are normalized cardinality rates.}
    \label{tab:clean_per_k_3percent}
    \begin{adjustbox}{width=0.95\columnwidth}
    \begin{tabular}{llrrrr}
        \toprule
        Method & $N$ & \downg{OSPA (m)} & \downg{mLE (m)} & \downg{MDR} & \downg{FAR} \\
        \midrule
        SymNetPro & 1 & \textbf{3.40} & 2.92 & \textbf{0.000} & \textbf{0.053} \\
        SymNetPro & 2 & \textbf{4.54} & 3.40 & \textbf{0.014} & 0.054 \\
        SymNetPro & 3 & \textbf{5.35} & 4.04 & \textbf{0.027} & \textbf{0.037} \\
        SymNetPro & 4 & \textbf{6.01} & 4.45 & \textbf{0.045} & 0.021 \\
        SymNetPro & 5 & \textbf{6.74} & 4.76 & 0.066 & 0.012 \\
        \midrule
        DeepMTL & 1 & 6.35 & \textbf{2.51} & 0.172 & 0.062 \\
        DeepMTL & 2 & 7.46 & \textbf{2.56} & 0.159 & \textbf{0.052} \\
        DeepMTL & 3 & 8.56 & \textbf{2.71} & 0.167 & 0.045 \\
        DeepMTL & 4 & 9.55 & \textbf{2.83} & 0.187 & 0.036 \\
        DeepMTL & 5 & 10.53 & \textbf{2.95} & 0.219 & 0.025 \\
        \midrule
        LRM-MSL & 1 & 11.69 & 8.14 & 0.075 & 0.266 \\
        LRM-MSL & 2 & 13.03 & 8.26 & 0.066 & 0.254 \\
        LRM-MSL & 3 & 13.47 & 8.25 & 0.059 & 0.243 \\
        LRM-MSL & 4 & 13.74 & 8.30 & 0.052 & 0.241 \\
        LRM-MSL & 5 & 13.90 & 8.34 & \textbf{0.047} & 0.235 \\
        \midrule
        cGAN & 1 & 11.16 & 18.20 & 0.005 & 0.191 \\
        cGAN & 2 & 13.58 & 21.97 & 0.151 & 0.107 \\
        cGAN & 3 & 14.80 & 24.39 & 0.265 & 0.052 \\
        cGAN & 4 & 15.73 & 24.98 & 0.376 & \textbf{0.020} \\
        cGAN & 5 & 16.43 & 24.24 & 0.466 & \textbf{0.007} \\
        \bottomrule
    \end{tabular}
    \end{adjustbox}
\end{table}

Table~\ref{tab:clean_per_k_3percent} reports localization performance as the number of active transmitters increases. SymNetPro shows the most stable OSPA degradation, increasing from 3.40~m for one transmitter to 6.74~m for five. DeepMTL achieves the lowest matched localization error but underestimates source cardinality increasingly often; at $N=5$, its average missed-source count is 1.095, corresponding to MDR $=0.219$. Thus, low mLE alone can be misleading because a method may accurately localize only a subset of easy sources. LRM-MSL shows the opposite tendency, with comparatively low MDR but high FAR, while cGAN increasingly under-detects sources. OSPA captures both localization and cardinality mismatch and therefore better reflects overall multi-source recovery quality.

\subsection{RQ5: Ablation Study}
\begin{figure*}
    \centering
    \includegraphics[width=1\linewidth]{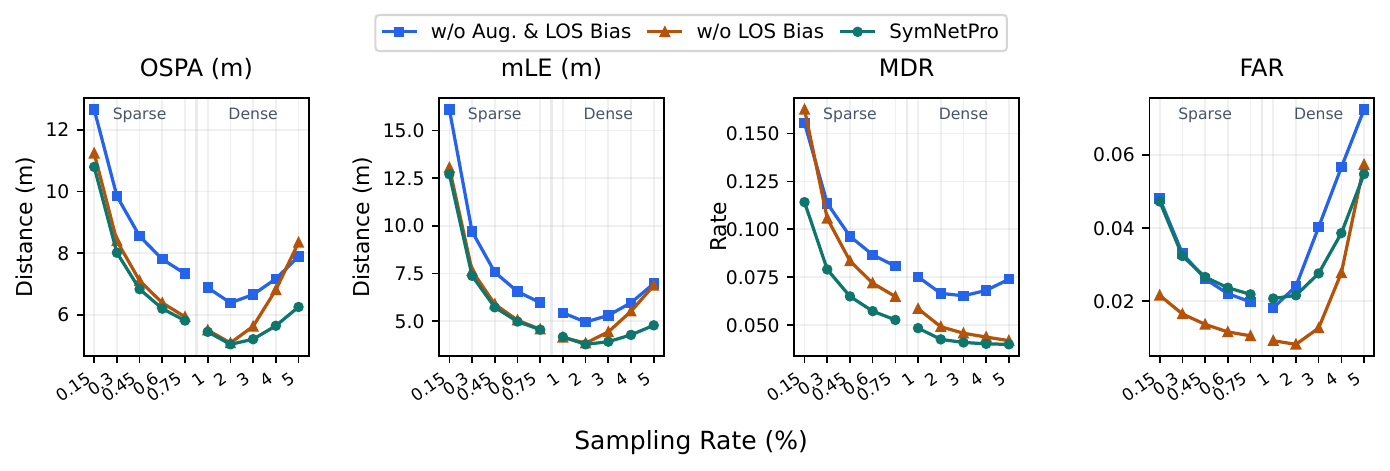}
    \Description{Ablation curves comparing the full model, the model without attention bias, and the model without both transmitter-drop and LOS bias across sampling densities.}
    \caption{Ablation study across sampling densities. In the plot, ``w/o LOS Bias'' removes the entire attention-bias injection while retaining transmitter-drop; ``w/o Aug. \& LOS Bias'' removes both additions.}
    \label{fig:abl}
\end{figure*}
We isolate the two proposed additions and two alternative bias designs. The \emph{w/o Bias} variant removes the entire attention-bias injection while retaining transmitter-drop. The \emph{w/o Aug. \& LOS Bias} variant removes both additions and is close to the original SymNet backbone~\cite{SymNet}. Variants without transmitter-drop are trained for twice as many epochs. Transmitter-drop gives the clearest gain under extreme sparsity, whereas the sparse geometric bias becomes more beneficial as the sensing density increases.

We additionally test a distance-only bias-score variant and a dense LOS-bias variant. The distance-only variant keeps the same sparse candidate graph but removes the NLOS penalty from the injected bias ($\lambda_{\mathrm{nlos}}=0$). The dense variant bypasses candidate-graph sparsification and applies the same LOS-aware bias to all valid patch pairs. In both alternatives, optimization stalled and the training loss showed little sustained decrease from its initial level. One possible explanation for the dense variant is that global pairwise bias introduces many weak or uninformative geometric preferences; however, the present experiment establishes the optimization failure rather than uniquely identifying its cause. The proposed sparse construction trained stably across the tested sampling range.

\subsection{Computational Efficiency}
\begin{table}[t]
\centering
\caption{End-to-end runtime and model size, averaged over 100 batches with batch size 64.}
\label{tab:efficiency}
\small
\begin{tabular}{lrrr}
\toprule
Method & Params & ms/batch & ms/sample \\
       & (M)    &          & (amort.) \\
\midrule
SymNetPro                 & 12.208 & 1294.62 & 20.23 \\
SymNetPro w/o bias        & 12.207 &  502.71 &  7.85 \\
cGAN+PH                   &  1.224 & 1972.09 & 30.81 \\
DeepMTL+YOLO-V3           & 61.576 &   89.21 &  1.39 \\
LRM-MSL+SourceNet         & 11.428 &   78.08 &  1.22 \\
\bottomrule
\end{tabular}
\end{table}
Runtime is measured on the same server with NVIDIA H100 80GB GPUs and dual Intel Xeon Platinum 8468 CPUs using 8 DataLoader workers. The timed region includes data loading and dataset-side preparation, model forward execution, and method-specific decoding/post-processing, while excluding metric computation. CPU-side post-processing is parallelized when applicable. SymNetPro timing includes online candidate-graph and bias construction from the precomputed LOS table.

The LOS bias changes the parameter count only from 12.207M to 12.208M, but increases amortized end-to-end runtime from 7.85 to 20.23~ms/sample. SymNetPro remains faster than cGAN+PH (30.81~ms/sample) but slower than DeepMTL+YOLO-V3 and LRM-MSL+SourceNet. The lower latency of the latter two should be interpreted together with sparse-regime recovery quality: for five-source scenes at 100 samples, SymNetPro misses 0.88 sources and produces 0.10 false detections on average, compared with 1.74 missed sources for DeepMTL and 4.09 false detections for LRM-MSL. SymNetPro therefore trades additional computation for substantially more reliable source recovery under severely limited sensing.

\subsection{RQ6: Failure Analysis with Exclusive Evidence}
\begin{figure}
    \centering
    \includegraphics[width=1\linewidth]{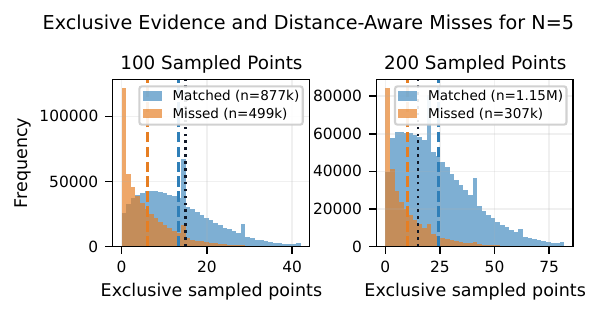}
    \Description{Distributions of transmitter-specific exclusive sample counts for matched and missed transmitters in five-source scenes at 100 and 200 sampled points.}
    \caption{Distribution of exclusive sampled points for matched and missed transmitters when $N=5$. Dashed lines mark the means. The dotted line at 15 is a worst-case joint-dimensionality reference ($3N$), not a hard per-transmitter detection threshold.}
    \label{fig:exclusive_samples}
\end{figure}

Figure~\ref{fig:exclusive_samples} analyzes transmitter-level failures for $N=5$ under 100 and 200 sampled points. Here, a transmitter is counted as missed only if it is not matched to any prediction within the 20~m localization threshold; this distance-aware definition is distinct from the cardinality-only MDR used in the main evaluation.

The dotted line at 15 equals the $3N$ continuous dimensionality of a five-transmitter directional configuration and is used only as a joint-problem reference scale; Appendix~\ref{app:joint} gives the corresponding rank-count argument and its limitations. When the other $N-1$ transmitters are not already constrained, recovering a target from aggregate observations can require jointly disambiguating their locations and orientations as nuisance parameters. The full joint inverse problem therefore contains $3N$ continuous unknowns, in addition to the discrete ambiguity of which transmitters are active. Importantly, the exclusive count $N_m(10)$ is \emph{not} the number of measurements in a full-rank joint Jacobian, so $N_m(10)\ge15$ is neither a necessary nor a sufficient per-transmitter identifiability condition. The reference simply places the target-specific evidence count against the dimensional scale of the unresolved five-source configuration.

The empirical distributions nevertheless show a strong association between exclusive evidence and misses. With 100 sampled points, matched transmitters have 13.4 exclusive samples on average, while missed transmitters have 6.0. The distance-aware miss rate is 45.6\% below the 15-sample reference and 12.2\% at or above it; 90.6\% of missed transmitters fall below 15. With 200 sampled points, matched and missed transmitters average 24.2 and 10.3 exclusive samples, respectively; the miss rate is 36.5\% below 15 and 9.3\% at or above it, and 74.9\% of misses remain below the reference. These results do not establish a deterministic threshold, but they show that localization failures are strongly concentrated in regimes with limited transmitter-specific evidence.

\section{Conclusion}
We presented SymNetPro for passive, source-unresolved directional multi-transmitter localization from sparse RSS. The method retains the SymNet dual-task reconstruction/localization backbone and adds two targeted extensions: a sparse LOS-aware attention bias and transmitter-drop augmentation. Experiments show substantially lower OSPA and more reliable source-cardinality recovery under sparse sensing, with consistent robustness to test-time measurement noise and increasing source count. The geometric prior introduces essentially no parameter growth but nontrivial computational overhead, yielding an explicit accuracy--efficiency trade-off in the target sparse regime.

Our failure analysis connects source superposition to transmitter-specific identifiability. Pairwise exclusive evidence provides a conservative indicator of observations in which a target contributes strongly to the aggregate field, while the SIR/sensitivity derivation explains why weak contributions are attenuated in the dB observation. Missed transmitters are strongly concentrated in low-evidence regimes, suggesting that further gains may require sensing strategies that actively increase transmitter-specific information. The current evaluation remains synthetic; validation on measured directional multi-transmitter data is an important next step.

\begin{acks}
Research was partially sponsored by the Army Research Laboratory and was accomplished under Cooperative Agreement Number W911NF-23-2-0014. The views and conclusions contained in this document are those of the authors and should not be interpreted as representing the official policies, either expressed or implied, of the Army Research Laboratory or the U.S. Government. The U.S. Government is authorized to reproduce and distribute reprints for Government purposes notwithstanding any copyright notation herein.
\end{acks}

\printbibliography
\appendix
\section{Additional Analysis and Implementation Details}
\label{app:theory}

\subsection{dB-domain Superposition and Conventional SIR}
\label{app:superposition}

Let $S_q(x)$ denote the linear received power from active transmitter $q$ at location $x$, and let
\begin{equation}
S_{\mathrm{agg}}(x)=\sum_{q\in\mathcal{T}}S_q(x),
\qquad
P_{\mathrm{agg}}^{\mathrm{dB}}(x)=10\log_{10}S_{\mathrm{agg}}(x).
\end{equation}
Because powers add in the linear domain, dB superposition can hide weak sources. Let $S_{(1)}(x)=\max_{q\in\mathcal{T}}S_q(x)$ denote the strongest component. Then
\begin{equation}
1\le \frac{S_{\mathrm{agg}}(x)}{S_{(1)}(x)}\le N,
\end{equation}
which gives the general dominance bound
\begin{equation}
0\le P_{\mathrm{agg}}^{\mathrm{dB}}(x)-10\log_{10}S_{(1)}(x)
\le 10\log_{10}N.
\end{equation}
For two equally strong sources the upper bound is $10\log_{10}2\approx3.01$~dB; with at most five sources it is $10\log_{10}5\approx6.99$~dB. These are upper bounds on the aggregate increase over the strongest component, not guarantees that every source leaves a visible dB signature.

For a particular target transmitter $m$ and $N\ge2$, define aggregate interference from the other active transmitters as
\begin{equation}
I_m(x)=\sum_{q\in\mathcal{T}\setminus\{m\}}S_q(x),
\end{equation}
and define the conventional linear and dB signal-to-interference ratios by
\begin{equation}
\rho_m(x)=\frac{S_m(x)}{I_m(x)},
\qquad
\mathrm{SIR}_m^{\mathrm{dB}}(x)=10\log_{10}\rho_m(x).
\end{equation}
Relative to the interference-only field, adding the target changes the aggregate dB observation by exactly
\begin{equation}
\Delta P_m(x)
=10\log_{10}\frac{S_m(x)+I_m(x)}{I_m(x)}
=10\log_{10}\!\left(1+\rho_m(x)\right).
\label{eq:delta_sir}
\end{equation}
Equivalently,
\begin{equation}
\Delta P_m(x)=10\log_{10}\!\left(1+10^{\mathrm{SIR}_m^{\mathrm{dB}}(x)/10}\right).
\end{equation}
Thus, at $\mathrm{SIR}=-10$~dB a target changes the aggregate measurement by only about $0.41$~dB, and at $-20$~dB by about $0.043$~dB. For scale, both changes are much smaller than the 4~dB test-noise standard deviation used in RQ3, although localization can still exploit multiple measurements and spatial structure rather than a single sample. Equation~\eqref{eq:delta_sir} is therefore a per-sample visibility argument, not a standalone detection bound.

\subsection{Local Sensitivity and Information Scaling}
\label{app:sensitivity}

The SIR interpretation can be connected directly to local sensitivity. At a sampled location $s_k$, let $\psi_m=(x_m,y_m,\phi_m)$ denote the target's directional parameters and suppose the local propagation response is differentiable at the configuration under consideration. The aggregate dB observation is
\begin{equation}
y_k=10\log_{10}\!\left(\sum_{q\in\mathcal{T}}S_q(s_k;\psi_q)\right).
\end{equation}
Differentiating with respect to $\psi_m$ gives
\begin{equation}
\nabla_{\psi_m}y_k
=\frac{10}{\ln 10}\,
\frac{S_m(s_k;\psi_m)}{\sum_{q\in\mathcal{T}}S_q(s_k;\psi_q)}
\nabla_{\psi_m}\ln S_m(s_k;\psi_m).
\end{equation}
Define the target's fractional contribution to the aggregate linear power as
\begin{equation}
\alpha_{k,m}
=\frac{S_m}{S_m+I_m}
=\frac{\rho_m}{1+\rho_m}
=\frac{1}{1+10^{-\mathrm{SIR}_m^{\mathrm{dB}}/10}}.
\label{eq:alpha_sir}
\end{equation}
Then
\begin{equation}
\nabla_{\psi_m}y_k
=\frac{10}{\ln 10}\,\alpha_{k,m}\,g_{k,m},
\qquad
g_{k,m}:=\nabla_{\psi_m}\ln S_m(s_k;\psi_m).
\label{eq:sensitivity_alpha}
\end{equation}
This makes the role of source superposition explicit: the geometry-dependent gradient $g_{k,m}$ is multiplicatively attenuated by the target's fractional power contribution $\alpha_{k,m}$.

The same factor appears quadratically in a local least-squares information measure. Let $J_{k,m}=\nabla_{\psi_m}y_k^{\top}$ be the $1\times3$ Jacobian row. Its Gauss--Newton curvature contribution is
\begin{equation}
J_{k,m}^{\top}J_{k,m}
=\left(\frac{10}{\ln 10}\right)^2
\alpha_{k,m}^{2}\,g_{k,m}g_{k,m}^{\top}.
\label{eq:gn_alpha}
\end{equation}
Under a local homoscedastic Gaussian dB-noise approximation with variance $\sigma_y^2$, the corresponding per-sample Fisher-information contribution has the same matrix up to the factor $1/\sigma_y^2$. Therefore, holding the propagation-gradient term fixed, reducing the target fraction by a factor of ten reduces this local information scale by a factor of one hundred.

High SIR is not by itself sufficient for localization. Equation~\eqref{eq:gn_alpha} also depends on the magnitude and direction of $g_{k,m}$: many high-SIR samples concentrated in similar geometric locations can contribute redundant Jacobian rows, while lower-SIR samples may still be useful when their gradients add complementary spatial information. This is why the evidence count used below is interpreted as a diagnostic proxy rather than an identifiability theorem.

\subsection{Pairwise Exclusive Evidence and Its Relation to Aggregate SIR}
\label{app:pairwise}

The empirical failure analysis was computed using a target-versus-strongest-competitor margin rather than conventional aggregate SIR. Define
\begin{equation}
D_m(s_k)
=P_m^{\mathrm{dB}}(s_k)
-\max_{q\in\mathcal{T}\setminus\{m\}}P_q^{\mathrm{dB}}(s_k).
\end{equation}
Let
\begin{equation}
M_m(s_k)=\max_{q\in\mathcal{T}\setminus\{m\}}S_q(s_k).
\end{equation}
Because
\begin{equation}
M_m(s_k)\le I_m(s_k)\le (N-1)M_m(s_k),
\end{equation}
the pairwise margin and conventional aggregate SIR satisfy the pointwise sandwich bound
\begin{equation}
D_m(s_k)-10\log_{10}(N-1)
\le \mathrm{SIR}_m^{\mathrm{dB}}(s_k)
\le D_m(s_k).
\label{eq:pairwise_sir_sandwich}
\end{equation}
Thus the two quantities are related but are not interchangeable. In particular, the exclusive-evidence criterion
\begin{equation}
D_m(s_k)>\delta
\end{equation}
implies
\begin{equation}
\mathrm{SIR}_m^{\mathrm{dB}}(s_k)
>\delta-10\log_{10}(N-1),
\end{equation}
and, using Equation~\eqref{eq:alpha_sir},
\begin{equation}
\alpha_{k,m}>
\frac{1}{1+(N-1)10^{-\delta/10}}.
\label{eq:pairwise_alpha_bound}
\end{equation}
For the reported $\delta=10$~dB and $N=5$, an exclusive sample therefore has aggregate $\mathrm{SIR}>3.98$~dB and $\alpha_{k,m}>0.714$. The 10~dB number in Figure~\ref{fig:exclusive_samples} is consequently a pairwise dominance threshold, not a 10~dB aggregate-SIR threshold.

The opposite regime is also informative. If the target is at least $\delta$~dB weaker than the strongest competitor, $D_m\le-\delta$, then $I_m\ge10^{\delta/10}S_m$ and hence
\begin{equation}
\mathrm{SIR}_m^{\mathrm{dB}}\le-\delta,
\qquad
\alpha_{k,m}\le\frac{1}{1+10^{\delta/10}}.
\end{equation}
At $\delta=10$~dB this yields $\alpha_{k,m}\le1/11\approx0.091$ and $\alpha_{k,m}^2\le0.0083$. Under comparable propagation gradients, Equation~\eqref{eq:gn_alpha} therefore assigns less than one percent of the local curvature scale of a target-dominated observation to such a strongly non-dominant sample. This provides the physical interpretation for using $N_m(10)$ as a conservative post-hoc indicator of strong transmitter-specific evidence.

\subsection{Joint versus Conditional Identifiability}
\label{app:joint}

The 15-sample marker in Figure~\ref{fig:exclusive_samples} is a dimensional reference for the unresolved joint inverse problem, not a theorem that an individual transmitter requires 15 exclusive samples. To make this distinction explicit, write the target parameters as $\psi_m\in\mathbb{R}^3$ and collect the remaining directional transmitters into
\begin{equation}
\eta_m=(\psi_q)_{q\in\mathcal{T}\setminus\{m\}}
\in\mathbb{R}^{3(N-1)}.
\end{equation}
Conditioned on a fixed active set of size $N$, the full continuous parameter vector is
\begin{equation}
\vartheta_m=(\psi_m,\eta_m)\in\mathbb{R}^{3N}.
\end{equation}
For $M$ scalar aggregate observations, let
\begin{equation}
F:\mathbb{R}^{3N}\rightarrow\mathbb{R}^{M},
\qquad
J=\frac{\partial F}{\partial\vartheta_m}
=\begin{bmatrix}J_m & J_{\eta}\end{bmatrix}.
\end{equation}
If the target and all nuisance transmitters must be jointly resolved from these observations, local full-column-rank identifiability of the continuous configuration requires
\begin{equation}
\operatorname{rank}(J)=3N,
\end{equation}
which is possible only if
\begin{equation}
M\ge3N.
\label{eq:joint_rank_count}
\end{equation}
For $N=5$, this dimensional lower bound is 15 scalar observations. Unknown source cardinality and the unknown active set add a separate discrete model-order ambiguity that is not counted by Equation~\eqref{eq:joint_rank_count}.

The conditional problem can be much smaller. If the nuisance configuration $\eta_m$ has already been determined by other observations or effectively constrained by prior information, recovering only $\psi_m$ concerns a three-dimensional continuous parameter block. Even then, three scalar measurements are not automatically sufficient: the rows of $J_m$ must be informative and linearly independent, and blockage or unfavorable geometry can make the local problem rank deficient. Conversely, joint recovery can succeed with fewer target-dominant samples because observations that are not counted by $N_m(10)$ can still constrain the nuisance transmitters or contribute complementary gradients.
\begin{table*}[h!]
\centering
\caption{Implementation details and hyperparameter settings.}
\label{tab:hyperparameters}
\small
\begin{adjustbox}{width=0.95\linewidth}
\begin{tabular}{ll@{\hspace{2.5em}}ll}
\toprule
\textbf{Category} & \textbf{Setting} &
\textbf{Category} & \textbf{Setting} \\
\midrule

Input channels &
Sparse signal map $R$, building/sampling mask $B_{\Omega}$ &
Radio-map loss &
MSE \\

Fixed RSS normalization &
$P_{\min}=-120$ dBm, $P_{\max}=-5.041252$ dBm &
Localization-map loss &
MSE on Gaussian transmitter\\&&&-location map \\

Patch size &
$8\times8$ &
Localization loss weight &
$1/2.5$ for full view; $1$ for drop view \\

Embedding dimension &
192 &
Total loss scaling &
$0.25(\mathcal{L}_{\mathrm{full}}+\mathcal{L}_{\mathrm{drop}})$ \\

Transformer depth &
6 &
Gaussian target std. &
$\sigma=17$ pixels \\

Attention heads &
16 &
Coordinate extraction &
Threshold 0.25 + NMS ($11\times11$ window) \\

MLP ratio &
4 &
Optimizer &
Adam; batch size 16 \\

Position embedding &
2D sine-cosine &
Base learning rate &
$5\times10^{-4}$ \\

Normalization &
Dynamic Tanh~\cite{tanh} &
Local-feature extractor and task-head LR &
$2.5\times10^{-4}$ \\

\midrule

LOS bias type &
Sparse patch-level attention bias &
Distance penalty $\lambda_d$ &
2.0 \\

Distance buckets &
$[0,50),[50,100),[100,\infty)$ pixels &
NLOS penalty $\lambda_{\mathrm{nlos}}$ &
1.0 \\

Bucket candidates &
$(12,12,8)$ near/mid/far &
Bias temperature $\tau_{\mathrm{bias}}$ &
2.0 \\

Patch confidence &
$q_i=\min(1,n_i/4)$ &
Initial bias scale $\gamma$ &
0.01 \\

Coarse distance weight $\lambda_c$ &
1.0 &
Final selected neighbors &
$K=32$ \\

Candidate LOS penalty $\lambda_{\mathrm{los}}$ &
1.0 &
LOS edge attribute &
$\ell_{ij}\in\{0,1\}$ \\

Candidate temperature $\tau_c$ &
0.12 &
& \\

\bottomrule
\end{tabular}
\end{adjustbox}
\end{table*}
\begin{figure*}[h!]
    \centering
    \includegraphics[width=1\linewidth]{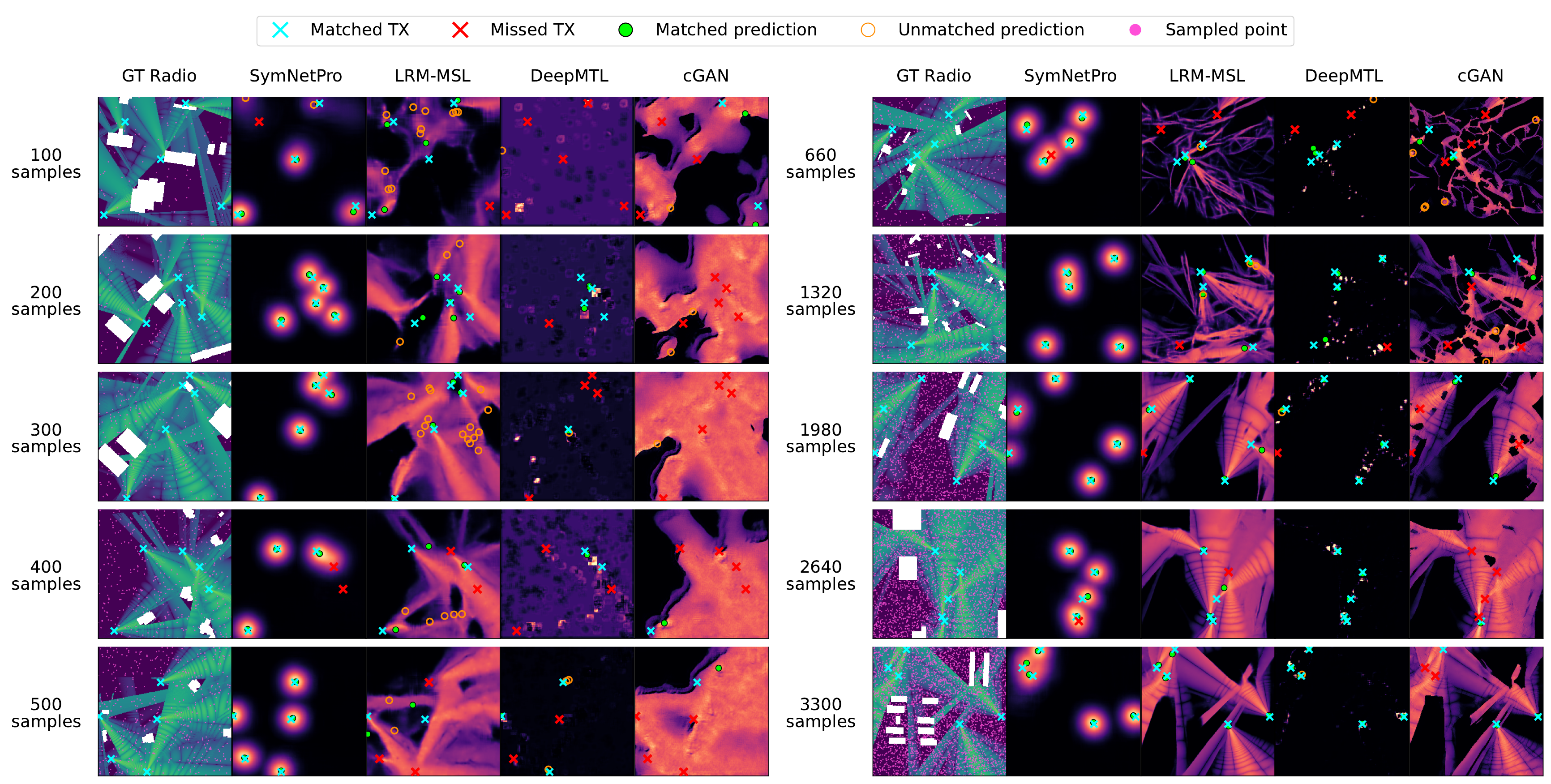}
    \Description{Qualitative localization outputs for SymNetPro and comparison methods over representative sparse and denser sensing settings.}
    \caption{Qualitative localization examples for the sparse and denser evaluation regimes.}
    \label{figf}
\end{figure*}
Therefore, the exclusive count $N_m(10)$ and the observation count $M$ in Equation~\eqref{eq:joint_rank_count} are fundamentally different quantities. The former counts only target-dominant sampled locations; the latter counts scalar observations available to the full joint inverse problem. We use the line at 15 only to place the empirical target-specific evidence count against the continuous dimensional scale of the unresolved five-source configuration. It is neither a necessary nor a sufficient per-transmitter threshold. If transmitter orientations were known rather than estimated, the analogous continuous joint dimension would be $2N$; this is likewise a dimensional comparison, not a universal sample-complexity law.

\subsection{Qualitative Localization Examples}
\label{app:qualitative}

The examples in Figure~\ref{figf} complement the aggregate metrics by showing the two dominant baseline failure modes under sparse observations: missed sources when only a subset of transmitters leaves strong recoverable evidence, and false alarms when reconstruction artifacts or fragmented directional responses are converted into extra source hypotheses. SymNetPro is not immune to evidence-limited cases, but its predictions remain more stable across the sparse-to-denser sensing range used in RQ1 and RQ2.

\label{app:implementation}

\end{document}